  \documentclass[twocolumn,pra,showpacs,floatfix]{revtex4}
\usepackage{graphicx}
\usepackage{subfigure}
\usepackage[latin1]{inputenc}
\usepackage[]{amsmath}
\usepackage{dcolumn}
\usepackage{bm}
\usepackage{tabularx}
\usepackage{color}

\def\ket#1{ $ \left\vert  #1   \right\rangle $ }  
\def\ketm#1{  \left\vert  #1   \right\rangle   }

\graphicspath{.}

\begin{document}

%
\title{MCDHF calculations of the electric dipole moment of radium 
       induced by the nuclear Schiff moment}
\author{Jacek Biero{\'n}}
\affiliation{Instytut Fizyki imienia~Mariana~Smoluchowskiego,
             Uniwersytet Jagiello{\'n}ski \\
             Reymonta~4, 30-059~Krak{ó}w, Poland \\
             \email{Bieron@uj.edu.pl}}
\author{Gediminas Gaigalas}
\affiliation{Vilnius University Research Institute of Theoretical Physics
             and Astronomy, \\
             A. Go\v{s}tauto 12, LT-01108 Vilnius, Lithuania}
\affiliation{Vilnius Pedagogical University,
            Student\c{u} 39, LT-08106 Vilnius, Lithuania}
\author{Erikas Gaidamauskas}
\affiliation{Vilnius University Research Institute of Theoretical Physics
             and Astronomy, \\
             A. Go\v{s}tauto 12, LT-01108 Vilnius, Lithuania}
\affiliation{Vilnius Pedagogical University,
            Student\c{u} 39, LT-08106 Vilnius, Lithuania}
\author{Stephan Fritzsche}
\affiliation{Helmholtzzentrum f{ü}r Schwerionenforschung (GSI), 
	     D-64291 Darmstadt, Germany}
\author{Paul Indelicato}
\affiliation{Laboratoire Kastler Brossel,
{É}cole Normale Sup{é}rieure \\
 CNRS; Universit{é} P.~et M.~Curie - Paris 6 \\
Case 74; 4, place Jussieu, 75252 Paris CEDEX 05, France}
\author{Per J{ö}nsson}
\affiliation{Nature, Environment, Society \\
             Malm{ö} University, S-205~06 Malm{ö}, Sweden}

\date{\today}

\begin{abstract}
The multiconfiguration Dirac-Hartree-Fock theory (MCDHF) has been
employed to calculate the electric dipole moment of the
$ 7s6d $~$ ^3 \! D _2 $
state of radium induced by the nuclear Schiff moment.
The results are dominated by
valence and core-valence electron correlation effects.
We show that the correlation effects  can be evaluated
in a converged series of multiconfiguration expansions.
\end{abstract}

\pacs{11.30.Er,21.10.Ky,24.80.+y,31.30.jg}













\maketitle

\section{Introduction}
\label{introduction}

A non-zero permanent electric dipole moment (EDM) of an atom, molecule, 
or any other composite or elementary particle is one of the possible 
manifestations of parity (P) and time reversal (T) symmetry violations. 
In the absence of any external electromagnetic field an atom can 
have a permanent EDM either due to an intrinsic EDM of one of its 
constituent particles or due to P-~and~T-violating (P-odd and T-odd) 
interactions between these 
particles~\cite{Martensson-Pendrill:1992,KhriplovichLamoreaux}.
When compared with the intrinsic EDM of the constituent particles, 
the net \textit{induced} EDM of an atom or molecule is often 
expected to be larger by several orders of magnitude due to
various nuclear and atomic enhancement mechanisms. Therefore, atoms and 
molecules are considered to be very attractive for carrying out EDM 
experiments and in the search for `new physics' beyond the standard model,
since, in the latter case, the induced EDM is greatly suppressed, when
compared to the anticipated values from the `new' 
theories~\cite{Ginges:2004}. In atomic physics, in particular, the experimental
search for a permanent EDM is gaining momentum due to recent 
advancements in trapping free neutral 
atoms~\cite{Legero:2004,Kuhr:2005,Beugnon:2006},
including various radioactive species~\cite{Guest:2007,Dammalapati:2007}.

During the last decade several atoms were considered as candidates 
for such experiments~\cite{Dzuba:2002,Ginges:2004}. These involved
(i) diamagnetic atoms (i.e., total angular momentum $J=0$)
in their respective ground states,
(ii) the alkalis, and (iii) atoms with a single $p$ electron 
outside closed shells, which were investigated in laser traps for their 
prospectives to perform EDM experiments.
Presently, radium appears to 
be the most promising candidate, and experiments on this element are under 
way at the Argonne National 
Laboratory~\cite{Scielzo:2006,Guest:2007,wwwArgonne} as well as the 
Kernfysisch Versneller Instituut~\cite{Jungmann:2002,Jungmann:priv,wwwKVI}.
The main advantages of radium lay in
(i) large nuclear charge $Z$,
(ii) simple electronic structure and closed-shell
[Kr]$ 4d^{10} 4f^{14} 5s^2 5p^6 5d^{10} 6s^2 6p^6 7s^2 $~$ ^1 \! S_0 $ 
ground state, 
(iii) octupole deformations of the radium nuclei for several
isotopes~\cite{Engel:2003,DobaczewskiEngel:2005}, as well as 
in (iv) coincidental proximity of two atomic levels of opposite parity,
$ 7s7p $~$ ^3 \! P _1 $ and
$ 7s6d $~$ ^3 \! D _2 $,
which are separated by a very small 
energy interval 5.41~cm$^{-1}$. In particular, the latter two advantages 
give rise to a relatively large enhancement factor, which is one of the 
largest among the atoms considered so far~\cite{Dzuba2000,Dzuba1999}.

However, extraction of fundamental P-~and T-violating parameters 
or coupling constants from experimentally measured atomic EDM requires 
atomic form factors which can be provided only by an {\sl ab initio} 
atomic theory.
Several of these form factors have been previously calculated
by the group
of Flambaum~\cite{Dzuba2000,Dzuba:2002}.
In practice, there are essentially four different form factors, 
related to four mechanisms which can induce an atomic EDM.
An atom can acquire a permanent EDM due to
P-~and~T-odd electron-nucleon interactions, or due to
the electromagnetic interaction of atomic electrons
with nuclear P-~and~T-odd moments, of which the leading ones are:
the Schiff moment, the magnetic quadrupole moment, and the
electric octupole moment, respectively. The latter two 
moments may exist only in nuclei with spins larger than~$I=1/2$, 
while the Schiff moment may exist also in isotopes with nuclear spin~$I=1/2$.
Nuclei with spins larger than~$I=1/2$ produce electric quadrupole 
shifts which are difficult to account for in an EDM measurement.
Therefore an EDM induced by the Schiff moment in~$I=1/2$ isotopes
seems to be the property of choice among (most of) the
experimenters~\cite{Jungmann:priv,Guest:priv,Trimble:priv}.
The nuclear Schiff moment is a P-odd and T-odd (electric-dipole) 
moment that occurs due to P- and T-violating interactions
at the nuclear scale. The Schiff
moment mixes atomic states of opposite parity and may induce static
EDM in atoms if magnetic and finite-size effects are taken into account
in the electron-nucleus interaction~\cite{Ginges:2004}. 

In the present paper, we present calculations for the
atomic EDM in radium, as induced by the Schiff moment.
The atomic wave functions were obtained within the framework of the 
multiconfiguration Dirac-Hartree-Fock (MCDHF) theory. The wave functions 
were separately optimized for the
$ 7s6d $~$ ^3 \! D _2 $ and
$ 7s7p $~$ ^3 \! P _1 $
states
(similar calculations were carried out recently for the 
scalar-pseudoscalar contribution to the EDM in cesium~\cite{Gaigalas2008}).
The main purpose of this paper is to provide a systematic 
evaluation of the effects of electron correlation on the calculated EDM 
of radium. We demonstrate the saturation of the
core-valence correlations, the dominant electron correlation effect beyond 
the  Dirac-Fock approximation.
%


\section{Theory}
\label{theory}

The Hartree-Fock and Dirac-Hartree-Fock theories, based either on the
finite-grid, basis-set, or some other numerical methods provide a 
natural point of departure in describing the electronic structure of atoms 
and molecules. For medium and heavy elements, these methods are often combined 
with Breit-Pauli or Dirac-Coulomb-Breit Hamiltonians in order to account 
for relativistic and retardation effects on the wave functions and the 
level structure of complex atoms. However, the main hindrance
in applying modern 
computational techniques arises from the electron-electron correlation, 
i.e., the residual interaction among the electrons beyond the atomic 
\textit{mean} field, and this is especially true for systems with many 
electrons. In neutral or nearly-neutral systems, missing electron-correlation 
effects are indeed often the main reason for the discrepancies 
between the observed and calculated properties of atoms. 

Today, many-body perturbation theory 
(MBPT)~\cite{LindgrenMorrison:86,Dzuba:CI+MBPT:2005,DzubaFlambaum:2007,%
JohnsonBook:2007} and various variational methods, often referred to as 
the multiconfiguration Hartree-Fock (MCHF) theory (or its relativistic 
counterpart --- the Dirac-Hartree-Fock 
theory~\cite{Grant:1988,GrantBook2007,grasp2K})
are the two dominant pillars in performing atomic structure calculations.
These methods are designed to evaluate the electron correlation effects
in a systematic manner.
For MBPT and most related methods, a nearby closed-shell 
configuration of the atom or ion is typically a convenient starting point. 
During the past decades, therefore, MBPT techniques were mainly applied to 
systems with either no or just a single electron outside closed 
shells. Systems with several electrons in open shells are, in contrast, much
more difficult to deal with if benefit is to be taken from the theory of 
angular momentum and spherical tensor operators, i.e., by
using a \textit{restricted} representation of the one-electron orbitals.
Difficulties occur then due to the large departure from a closed-shell 
$V^N$ potential and the rapid increase in the complexity of all 
perturbation expansions with any additional electron outside closed
shells. Although in the variational multiconfiguration methods
the algebraic complexity also depends on the shell structure of the atom, 
these methods can be applied more easily to systems with an arbitrary 
number of electrons outside closed shells. Apart from the number of
open shells, the accuracy of multiconfiguration Dirac-Hartree-Fock 
calculations depends crucially also on the occupation of the
valence shell(s) but this occurs rather indirectly, through the limiting 
number of configuration state functions that can be included in a 
particular wave function expansion. In this sense, the limitations are 
less conceptual but arise from the available computer resources. 
These limitations are typically related to the structure of the valence 
shell(s), i.e., to the angular properties of the valence electrons and 
their couplings.

\subsection{MCDHF theory}
\label{mcdhf}
We used a slightly modified version of the General Relativistic Atomic 
Structure Package (GRASP)~\cite{grasp2K} to generate the electronic 
wave functions.
In the multiconfiguration Dirac-Hartree-Fock method, the 
wave function for a particular atomic  state 
$ \Psi({\it \gamma} P J M_J ) $
is obtained as a linear combination of configuration state functions (CSFs) 
which are eigenfunctions of the parity $P$ and the total angular
momentum operators $J^2$ and $J_z$,
\begin{equation}
\label{ASF}
\Psi({\it \gamma} P J M_J ) = 
\sum_{r}^{\rm NCF} c_{r} \Phi(\gamma_{r} P J M_J ) \, .
\end{equation}
In the present computations, the wave functions 
were separately 
generated for the
$ 7s7p \;\, ^3 \! P _1 $ and
$ 7s6d \;\, ^3 \! D _2 $ states of radium.
Each wave function was obtained as self-consistent solution of the
Dirac-Hartree-Fock equations~\cite{Grant:1988} by using a
systematically increased multiconfiguration bases (of size $\rm NCF$) 
of symmetry-adapted configuration state functions 
$ \Phi(\gamma_{r} P J M_J ) $.
%
Configuration mixing coefficients $ c_{r} $ were obtained
through the diagonalisation of the Dirac-Coulomb Hamiltonian
\begin{equation}
\label{Dirac-Coulomb-Hamiltonian}
\hat{H}_{\rm DC} = \sum_{j=1}^{N} \left[c {\bm{ \alpha }}_j \cdot
                    {\bm{ p }}_j
         + (\beta_j -1)c^2 + V(r_j) \right]
         + \sum_{j>k} 1/r_{jk}
\end{equation}
where $V(r)$ is the monopole part of the electron-nucleus interaction.
A more detailed description of the
theory~\cite{GrantBook2007,Grant:1994}
and method of calculation~\cite{Bieron:Li:1996,Bieron:BeF:1999}
can be found elsewhere.

\begin{figure}
%
\resizebox{0.93\columnwidth}{!}%
{
 \includegraphics[height=6cm,width=8.6cm,angle=0]{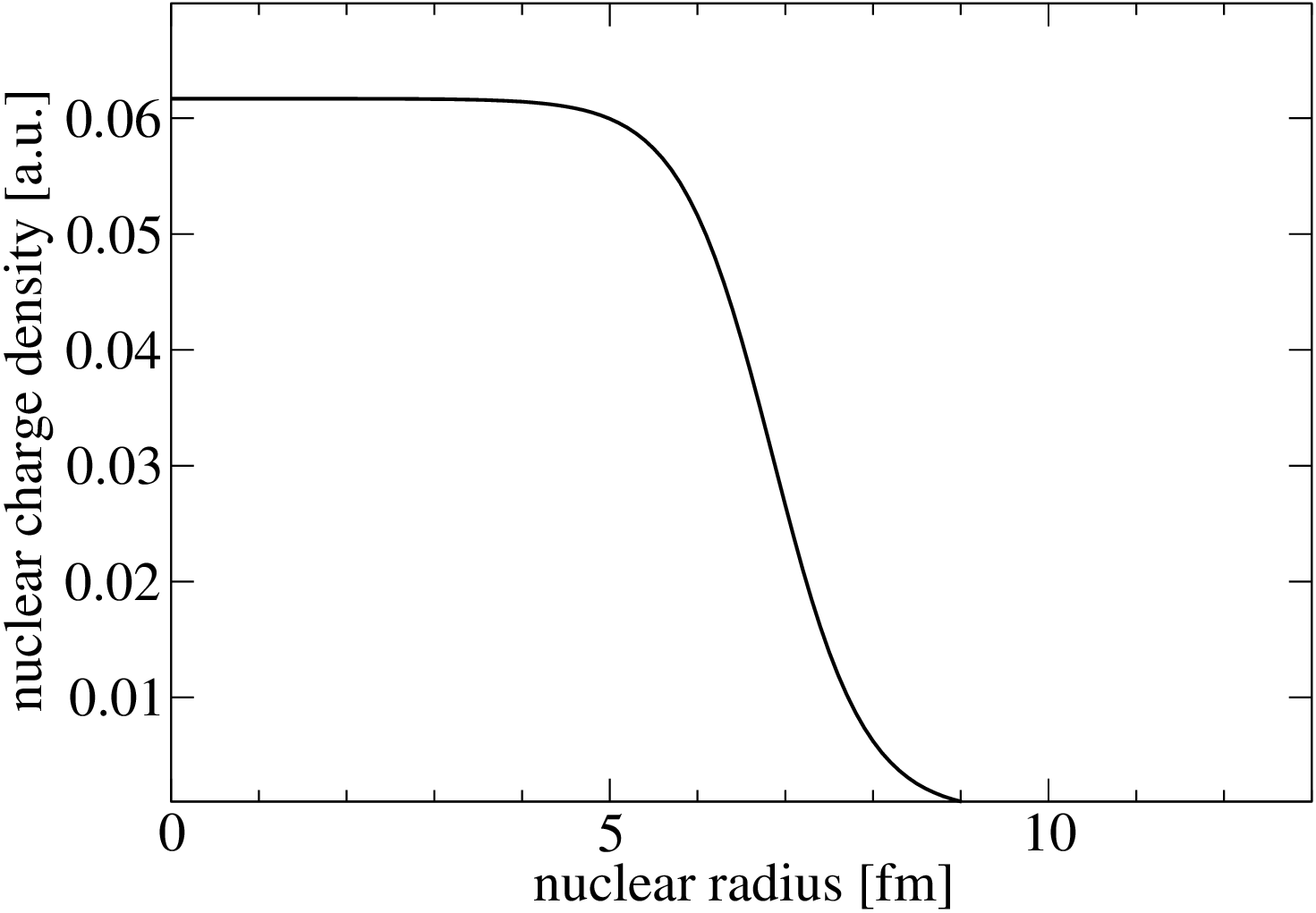}
}
\caption{Nuclear charge density $ \rho $ [a.u.] as function of the
nuclear radius $ r $ [fm], using a two-parameter Fermi distribution 
for the {${}^{225} _{\hspace{4pt}88}$Ra} nucleus.}
\label{fermi2p.fig}
\end{figure}

\subsection{Nuclear shape}
\label{grid}


All self-consistent-field calculations were done with the nucleus modelled
as a spherical ball by applying a two-parameter Fermi distribution 
\begin{eqnarray}
\label{two-parameter-fermi.eq}
   \rho(r) = \frac{\rho_0}{1 + \exp [(r-c) / a]}
\end{eqnarray}
in order to approximate the radial dependence of the nuclear charge 
density $ \rho(r) $. The parameter $\rho_0$ is derived from
the normalization condition $\int \rho(r) d^3r \,=\, Ze $.
Figure~\ref{fermi2p.fig} shows the nuclear charge density
$ \rho(r) $  inside the $ ^{225}_{\phantom{1}88}$Ra nucleus, calculated 
with the parameters 
$   c = 6.85            $~fm  and $ a = 0.523            $~fm  
%
%
%
(see~\cite{grasp89,JohnsonSoff:1985} for details).
For the other two isotopes the 'half-charge-density' parameter $c$ was set to
$   c = 6.83            $~fm ($ ^{223}_{\phantom{1}88}$Ra) 
and
$   c = 6.73            $~fm ($ ^{213}_{\phantom{1}88}$Ra),
with the 'nuclear-skin-thickness' $a$ value unchanged.
These nuclear charge distributions were also used in the subsequent
calculations of the Schiff moment expectation
values~(Eq.~\ref{eq:radial_integral_Schiff}).

In the GRASP code~\cite{grasp2K}
 all electronic (radial) orbitals are represented
on a numerical grid which increases exponentially
in order to ensure an accurate representation
of the atomic wave functions near the nucleus.
The grid is generated from the formula
$r_i \,=\, r_o \exp ((i-1)h)$,
with
$ r_o \,=\, 2.0 × 10^{-8} × a_0 $, \ 
$ h   \,=\, 7.0 × 10^{-3}     $, and  \ 
$ i \,=\, 1 ,\ldots, 4000$.
With the above parameters,
there were 1255 grid points within the 'half-charge-density' nuclear radius
$r_{\rm nuc} \,=\, 6.85 $~fm,
while the numerical representation of the full extent of all radial 
(core and valence) electronic orbitals required more than 3000 points.
With the above choice of the radial grid, all necessary one- and
two-particle matrix elements can be calculated with a (relative) accuracy
of the order of $\sim 10^{-8}$ or better.
Note however that larger uncertainties may arise
from the radial matrix elements in Eq.~(\ref{eq:radial_integral_Schiff}),
due to the approximate nature of the nuclear charge
distribution in Eq.~(\ref{two-parameter-fermi.eq})
and due to deviations from radial symmetry of those isotopes, 
for which nuclear deformations are significant
(for model dependence see~\cite{www.FiniteNuclei}).

\subsection{Atomic EDM}
\label{edm}
Neglecting the contributions from the off-diagonal hyperfine interaction 
the coupled wave function of the total system `electrons + nucleus' 
is given by the Clebsch-Gordan expansion~\cite{non-diagonal}
\begin{eqnarray}
\label{eq:Wave_function}
   \Psi  \left( \gamma \nu P J I F M_F \right) & = & 
   \nonumber \\ [0.2cm]
   &  & \hspace*{-3.5cm}
   \sum_{M_J M_I}
   \left< J I M_J M_I | J I F M_F \right>  \;
   \Psi \left( \gamma P J M_J \right) \; \Psi \left( \nu I M_I \right) \, 
\end{eqnarray}
where $\Psi \left( \nu I M_I \right)$ represents the ground state of the 
nucleus, and where the standard notation is used for the Clebsch-Gordan
coefficients. For high-$Z$ elements with closed levels of opposite parity,
such as radium in the $ ^3 \! P_1 $ and $ ^3 \! D_2 $ levels,
one of the most important 
parity (P) and time reversal symmetry (T) violating interactions is
caused by the nuclear Schiff moment $\mathbf{S}$,
which gives rise to the
electron-nucleus interaction:
\begin{eqnarray}
\label{eq:H_SM}
   \hat{H}_{SM}  \; = \;
   4 \pi \;
   \sum_{j=1} ^{N} \; \left( \bm{ S}  \cdot \bm{ \nabla}_{j} \right)
   \; \rho \left( r_j \right) \, .
\end{eqnarray}
In this Hamiltonian, $\rho \left( r \right)$ is the 
normalized to unity nuclear 
density function from Eq.~(\ref{two-parameter-fermi.eq}),
and the Schiff moment $\bm{S}$ is directed along the 
nuclear spin $\bm{I}$: $\bm{S} \, \equiv \, S \bm{I} / I$. The interaction
in Eq.~(\ref{eq:H_SM}) 
mixes states of different parity and may also induce a static electric 
dipole moment of the atom. Since the Schiff moment interaction is quite weak, 
we can express the wave function of the (mixed-parity) hyperfine state
\ket{F,M_F} of the level $J^P$ as~\cite{Gaigalas2008}:
\begin{widetext}
\begin{eqnarray}
\label{eq:mixed_parity}
   \widetilde{\Psi}  \left( \gamma \nu J I F M_F \right) \; & = &
   a \Psi  \left( \gamma \nu P J I F M_F \right) \; + \;
   \sum_{i=1} ^{m} \; b_i \; 
   \Psi \left( \alpha _i \nu (-P) J_i I F M_F \right) \, 
\end{eqnarray}
where the coefficient $a$ of the given hyperfine state can be 
set to 1. The expansion coefficients of the other (hyperfine) states of 
  opposite parity can be perturbatively approximated by
%
%
\begin{eqnarray}
\label{eq:b_factor}
%
   b_i \; = \;
   \frac{
   \left< \Psi \left( \alpha_i \nu (-P) J_i \; I F M_F \right) | \hat{H}_{SM}
    |\Psi  \left( \gamma \nu P J IF M_F \right) \right>}
   {E \left( \gamma P J \right) \; - \; E \left( \alpha_i (-P) J_i \right)} 
   \, .
\end{eqnarray}
The mixed-parity wave function in Eq.~(\ref{eq:mixed_parity})
for the hyperfine state \ket{F,M_F} of
a particular atomic level $^{2S+1}L_J$
induces a static EDM of an atom:
\begin{eqnarray}
\label{eq:DA}
   D_A  = 
   \left< \widetilde{\Psi}  \left( \gamma \nu J I F M_F \right) | \hat{D}_{z}
   | \widetilde{\Psi}  \left( \gamma \nu J I F M_F \right) \right> 
   \; = \; 
   \; 2  \sum_{i=1}^{m} \; b_i
   \left< \Psi \left( \gamma \nu P J I F M_F \right) | \hat{D}_{z} |
          \Psi \left( \alpha _i \nu (-P) J_i I F M_F \right)
   \right>
\end{eqnarray}
where $\hat{D}_z$ denotes the $z$ projection of the electric-dipole 
moment operator. For this electric-dipole operator, the matrix element between 
(hyperfine) states of different parity can be expressed as:
%
\begin{eqnarray}
\label{eq:mat_element_EDM}
   \left< \Psi \left( \gamma \nu P J IF M_F \right) | \hat{D}_{z}
    | \Psi \left( \alpha_i \nu (-P) J_i I F M_F \right) \right> & = &
   \nonumber  \\ [1ex]
   &  & \hspace*{-8.9cm}
   (-1)^{I+J+F+1} \;
   \left( 2F+1 \right) \; \sqrt{2J+1} \;
   \left(
   \begin{array}{ccc}
      F    & 1 & F \\
      -M_F & 0 & M_F
   \end{array}
   \right)  
%
   \left\{
   \begin{array}{ccc}
      J & F & I \\
      F & J_i & 1
   \end{array}
   \right\}
   \left[ \Psi  \left( \gamma P J \right) \| \hat{D}^1 \| 
          \Psi  \left( \alpha_i (-P) J_i \right) \right] \, 
\end{eqnarray}
while the matrix element of the (scalar) electron-nucleus interaction 
in Eq.~(\ref{eq:H_SM}), induced by the Schiff moment, is written as:
\begin{eqnarray}
\label{eq:mat_element_Schiff}
   \left< \Psi  \left( \gamma \nu P J IF M_F \right) | \hat{H}_{SM}
    | \Psi \left( \alpha_i \nu (-P) J_i \; I F M_F \right) \right> & = &
   \nonumber \\ [1ex]
   &  &  \hspace*{-8.9cm}
   (-1)^{I+J+F+1} \; 
   \sqrt{2J+1} \;
   \sqrt{ \frac{\left(I + 1 \right)\left(2I + 1 \right)}{I}}
   \left\{
   \begin{array}{ccc}
      I   & I & 1 \\
      J_i & J & F 
   \end{array}
   \right\}
%
  4\pi
   \; S \;
   \left[ \Psi \left( \gamma P J \right) \| 
          \hat{\nabla}^{1} \; \rho \left( r \right) \| 
	  \Psi \left( \alpha_i (-P) J_i \right) \right] 
 \hspace*{.4cm}
\end{eqnarray}
\end{widetext}
and is independent of $M_F$. In the following, we shall refer to the last 
term on the right-hand side of Eq.~(\ref{eq:mat_element_Schiff}) as the 
reduced matrix element of the Schiff operator for the two (fine-structure) 
levels $J$ and $J_i$ of different parity, and shall assume
$M_F \; = \; F$, in line with optical pumping schemes of hyperfine levels
with circularly polarized light.
For the \ket{F,M_F} hyperfine state of the $ ^3 \! D_2 $ level, the static
EDM in Eq.~(\ref{eq:DA}), induced by the P-odd and T-odd nuclear Schiff
moment, becomes
\begin{widetext}
\begin{eqnarray}
\label{eq:DA-3D2}
   D_A (^3 \! D_2,\, F M_F) \; = \;
%
   2 \; 
   \frac{ \left< \Psi  \left( ^3 \! D_2,\, F M_F \right) | 
                \hat{D}_{z} | 
		 \Psi \left( ^3 \! P_1,\, F M_F \right) \right> \:
          \left< \Psi \left( ^3 \! P_1,\, F M_F \right) | 
                \hat{H}_{SM} |  
		 \Psi  \left( ^3 \! D_2,\, F M_F \right) \right> }{
	 E \left( ^3 \! D_2 \right) \; - \; E \left( ^3 \! P_1\right)} \,
\end{eqnarray}
\end{widetext}
if the summation over the intermediate states is restricted 
to the nearby $ ^3 \! P_1 $ level.
This assumption is justified by the size of the energy denominator
which is 500~times smaller than the second smallest.
Moreover, the electron nucleus 
interaction in equation~(\ref{eq:H_SM})
is scalar, therefore only one of the hyperfine states may occur.

When the multiconfiguration expansion from Eq.~(\ref{ASF}) is employed
for the electronic 
part of the (total) wave functions, the reduced matrix elements of general 
tensor operator $\hat{T}^{k}_{q}$ can be decomposed
into (reduced) matrix elements between configuration state functions
\begin{eqnarray}
\label{eq:mat_element_between_ASF}
   \left[ \Psi  \left( \gamma P J \right) \| \hat{T}^{k} \| 
          \Psi  \left( \alpha (-P) J_i \right) \right]
   & = &
   \nonumber \\ [0.2cm]
   &  & \hspace*{-4.5cm}
 \sum_{r,s} \; c_r c_s
   \left[ \Phi  \left( \gamma_r P J \right) \| \hat{T}^{k} \| 
          \Phi \left( \gamma_s (-P) J_i \right) \right] \, 
\end{eqnarray}
and those, in turn, into a sum of single-particle matrix elements
\begin{eqnarray}
\label{eq:mat_element_between_CSF}
   \left[ \Phi \left( \gamma_r P J \right) \| \hat{T}^{k} \| 
          \Phi \left( \gamma_s (-P) J_i \right) \right] & = &
   \nonumber \\ [0.2cm]
   &  & \hspace*{-4.5cm}
   \; \sum_{a,b} \; d^{k}_{ab}(rs)  
   \left[ n_a \kappa_a \| \hat{t}^{k} \| n_b \kappa_b \right].
\end{eqnarray}
In the latter expansion, the $d^{k}_{ab}(rs)$ are known as 
`angular coefficients' that arise from using Racah's algebra in the 
decomposition of the many-electron matrix elements~\cite{GrantBook2007,ANCO2}.
The single-particle reduced matrix elements in the expansion 
(\ref{eq:mat_element_between_CSF}) can be factorized into reduced angular 
matrix elements and radial integrals which, for the Schiff moment 
interaction, read 
%
\begin{eqnarray}
\label{eq:radial_integral_Schiff}
   \left[ n_a \kappa_a \| 
   \hat{\nabla}^{1} \; \rho \left( r \right) \| n_b \kappa_b \right]
    & = & 
   \nonumber \\ [0.2cm]
   &  & \hspace*{-3.5cm}
   \left[ \kappa_a \| C^{1} \| \kappa_b \right] \;
   \int_{0}^{\infty}
   \left( P_a P_b \; + \; Q_a Q_b \right) \; \frac{d \rho}{dr} \; dr \, ,
\end{eqnarray}
while for the electric-dipole moment operator
($k \; = \; 1$) it is
\begin{eqnarray}
\label{eq:radial_integral_EDM}
   \left[ n_a \kappa_a \| \hat{d}^1 \| n_b \kappa_b \right]
    & = & 
   \nonumber \\ [0.2cm]
   &  & \hspace*{-2.9cm}
- \left[ \kappa_a \| C^{1} \| \kappa_b \right] \;
   \int_{0}^{\infty}
   \left( P_a P_b \; + \; Q_a Q_b \right) \; r \; dr.
\end{eqnarray}
For the calculations of the matrix elements we extended the 
GRASP~\cite{grasp2K} and
\emph{mdfgme}~\cite{mcdfgme} 
relativistic atomic structure packages. The extension, 
presented in this work, includes programs for both Schiff moment interaction 
and electric-dipole moment matrix elements.
Experimental energy differences were used in the calculations
of all expectation values.
The energy values for the two levels of interest
 ($ E_{7s6d \; {}^3 \! D_2} \; = \ 13993.97 cm^{-1} $ and
  $ E_{7s7p \; {}^3 \! P_1} \; = \ 13999.38 cm^{-1} $)
were taken from the tables of Moore~\cite{Moore:1971}.
The nuclear spin and magnetic moment data were taken from the tables of
Raghavan~\cite{Raghavan:1989}.

\subsection{Handling non-orthogonalities}
\label{biotra}

The electronic wave functions were optimised
separately for the two levels of interest.
All expectation values were evaluated with the biorthogonal technique
developed by Malmquist~\cite{Malmqvist1986}.
For two atomic state functions
\begin{equation}
   \Psi(\gamma PJ) = \sum_r c_r \Phi(\gamma_r PJ)
\end{equation}
and 
\begin{equation}
   \Psi(\alpha(-P)J_i) = \sum_s c_s \Phi(\gamma_s(-P)J_i) 
\end{equation}
the reduction of a general matrix element
\begin{equation}
   \left[ \Psi \left( \gamma P J \right) \| \hat{T}^{k} \| 
          \Psi \left( \alpha (-P) J_i \right) \right]
\end{equation}
into a sum of \textit{radial integral $×$ angular coefficient} terms
is based on tensor algebra techniques.
In the decomposition (\ref{eq:mat_element_between_CSF}), it is usually 
assumed that the (many-electron) configuration states on both sides of the
matrix element are built from a common set of spin-orbitals. This is a 
very severe restriction since a high-quality wave function demands orbitals 
optimized for the specific electronic state. Instead of 
the standard decomposition (\ref{eq:mat_element_between_CSF}) based on 
tensor algebra techniques, Malmquist has shown~\cite{Malmqvist1986} 
that for very general expansions, where the two atomic states are described 
by different orbital sets, it is possible to transform
the wave function 
representations of the two states in such a way that standard techniques
can be used for the reduction of the matrix elements in the new 
representation. This procedure has been implemented in the modules that 
compute Schiff moments and it can be summarized as follows
\begin{enumerate}
\item Perform MCDHF or CI calculations for the two states where the 
orbital sets of the two wave functions are not required to be identical.
\vspace*{-0.20cm}
\item Change the wave function representations by transforming the two 
orbital sets to a biorthogonal basis. This is followed by a 
counter-transformation of the expansion coefficients $c_r$ and $c_s$
so as to leave the resultant wave functions invariant.
\vspace*{-0.20cm}
\item
Calculate the matrix elements with the transformed wave functions for 
which now standard techniques can be used~\cite{GrantBook2007}.
\end{enumerate}  
The transformation of wave functions is very fast, since
it relies only on angular coefficients 
for a one-electron operator of rank zero which appears in the evaluation
of the kinetic energy term in the MCDHF or CI step.
The details of the transformations are discussed in~\cite{Olsen1995}.

%
\section{Method of calculation}
\label{method}
To generate the atomic states of interest, the method described as 
systematic expansion of configuration 
set \cite{Bieron:Li:1996,Bieron:BeF:1999} has been employed, in which
symmetry-adapted CSF of a given parity and total angular momentum are
generated by substitutions from reference configurations to an active 
set of orbitals. The active set should hereby comprise (at least) all 
valence shells, several `near-valence' core shells, as well as certain 
number of virtual shells. 
The active set and multiconfiguration expansions are then systematically 
increased until the expectation value (of interest) is converged. 
In practice, we divided the computations into three phases to
generate (i) the spectroscopic orbitals, (ii) the virtual orbitals, and
(iii) to perform large configuration interaction calculations, once 
the set of orbitals is fixed.

\subsection{Orbital set}
\label{orbitals}

%
%
In the first phase of the computations, all spectroscopic orbitals 
required to form a reference wave function were obtained with a minimal 
configuration expansion, with full relaxation. These orbitals 
were determined from a symmetry-adapted
Dirac-Fock calculation with only those configurations
which arise in $j$-$j$ coupling for a particular state of interest.
The spectroscopic orbitals were kept frozen in all later steps.
%

%
In the second phase the virtual orbitals were generated in five consecutive 
steps. At each step the virtual set has been extended by one layer
of virtual orbitals. A layer is defined as a set of virtual orbitals
with different angular symmetries. In the present paper
five layers of virtual orbitals were generated, each layer comprising
orbitals with symmetries {\sl s,p,d,f,g,} and {\sl h}.
At each step the configuration expansions were limited 
to single and double substitutions from valence shells to all new orbitals
as well as to all (virtual) orbitals of the previously generated 
layers. These substitutions were augmented by a small subset 
of dominant single and double substitutions from core and valence shells, 
with the further restriction that at most one electron may be
promoted from the core shells (i.e. in a double substitution
at least one electron is promoted from a valence shell).
All configurations from earlier steps were retained, with all previously 
generated orbitals fixed, and all new orbitals made orthogonal to 
all others of the same symmetry.
%
%
The initial shapes of radial orbitals were obtained by means of a
Thomas-Fermi potential, and then driven to convergence with
the self-consistency threshold set to
$10^{-10}$ for spectroscopic orbitals and
$10^{-8}$ for virtual orbitals, respectively.
All radial orbitals were separately optimized for each atomic state.
The Optimal Level form of the variational expression~\cite{grasp89}
was applied in all variational calculations.

\subsection{Configuration-interaction calculations}
\label{cic}

%
%
In the third phase of the computations, the confi­guration-interaction 
calculations (i.e.~without changing the radial shapes
of the one-electron spin-orbitals) were performed, with multiconfiguration 
expansions tailored in such a way, as to capture the dominant electron
correlation contributions to the expectation values.
All single and double substitutions were allowed from several core
shells and from both valence subshells (i.e.~$ 7s7p $, or $ 7s6d $, 
depending on the state) to all virtual shells, with the same restriction
as above, i.e.~that at most one electron may be promoted from core shells.
The virtual set was systematically increased from one to five layers.
In a similar manner, several core subshells were systematically
opened for electron substitutions --- from the outermost $6p$
up to the $5s 5p 5d 6s 6p$ subshells. 

The convergence of the calculations can be observed by monitoring the 
dependence of the 
matrix elements on the size of the virtual set as well as on the 
number of core subshells that are opened for electron 
substitutions.
The effects of substitutions from $4s4p4d4f$
and still deeper shells were estimated in our previous papers and turned
out to be below 1~percent in the case of hyperfine
structures~\cite{Bieron:RaQ:2005,Bieron:Rahfs:2005}
and a fraction of a percent in the case of transition
rates~\cite{Bieron:Ra3d2:2004,Bieron:Ratau:2007}.

%
\begin{table}
\caption{The values of the reduced matrix element of the Schiff operator
$4 \sqrt{3} \, \pi \, \left[ {}^3 \! P_1 \left\| \hat{\nabla}^{1} \;
    \rho \left( r  \right) \right\| {}^3 \! D_2 \right]$ 
from Eq.~(\ref{eq:mat_element_Schiff}) [a.u.].
Electron substitutions $from$ different sets of spectroscopic orbitals:
$      6p[7s6d|7s7p] $ (second column), 
$     6sp[7s6d|7s7p] $ (third column), ... ,
$ 5spd6sp[7s6d|7s7p] $ (sixth column).
Electron substitutions $to$ different sets of virtual orbitals:
 (1v) one layer of virtual orbitals,
 (2v) two layers of virtual orbitals, ... ,
 (5v) five layers of virtual orbitals.
'DF' = uncorrelated Dirac-Fock value.
}
\vspace*{0.3cm}
\label{Two_matrix_electron_Shiff_element}
\begin{tabular}{c|ccccccccc}
\colrule   \\[-0.3cm] 
 DF & \phantom{1}674 &  ~~&  ~~&  ~~& ~~&  \\ %
\colrule   \\[-0.3cm] 
\multicolumn{1}{c|}{ virtual } &
\multicolumn{5}{c}{ shells opened for substitutions  } & \\
\multicolumn{1}{c|}{ set } &
\multicolumn{1}{c}{ ~~~6p~~~  } &
\multicolumn{1}{c}{ ~~~6sp~~~ } &
\multicolumn{1}{c}{ ~~5d6sp~~ } &
\multicolumn{1}{c}{ ~5pd6sp~  } &
\multicolumn{1}{c}{ ~5spd6sp~ } \\
%
\colrule   \\[-0.3cm] 
 1v & $1931$ & $1892$ & $1862$ & $1749$ & $1647$ \\ %
 2v & $4597$ & $7544$ & $7553$ & $7598$ & $7754$ \\ %
 3v & $5141$ & $8073$ & $8112$ & $8125$ & $8327$ \\ %
 4v & $4760$ & $7936$ & $7947$ & $7889$ & $8117$ \\ %
 5v & $4642$ & $7834$ & $7818$ & $7743$ & $8217$ \\ %
\colrule  
%
\end{tabular}
\end{table}

\section{Results}
\label{results}

%
\begin{figure*}%
    \subfigure[ \ Schiff ]{
       \label{subfigure2_schiff}
       \includegraphics[width=.42\textwidth,height=.26\textheight]
         {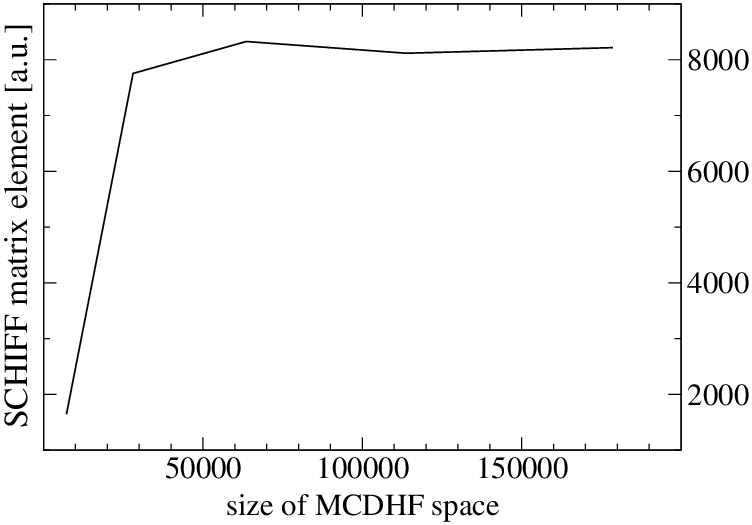}}
   \hspace{.5in}
    \subfigure[ \ EDM ]{
       \label{subfigure2_edm}
       \includegraphics[width=.42\textwidth,height=.26\textheight]
         {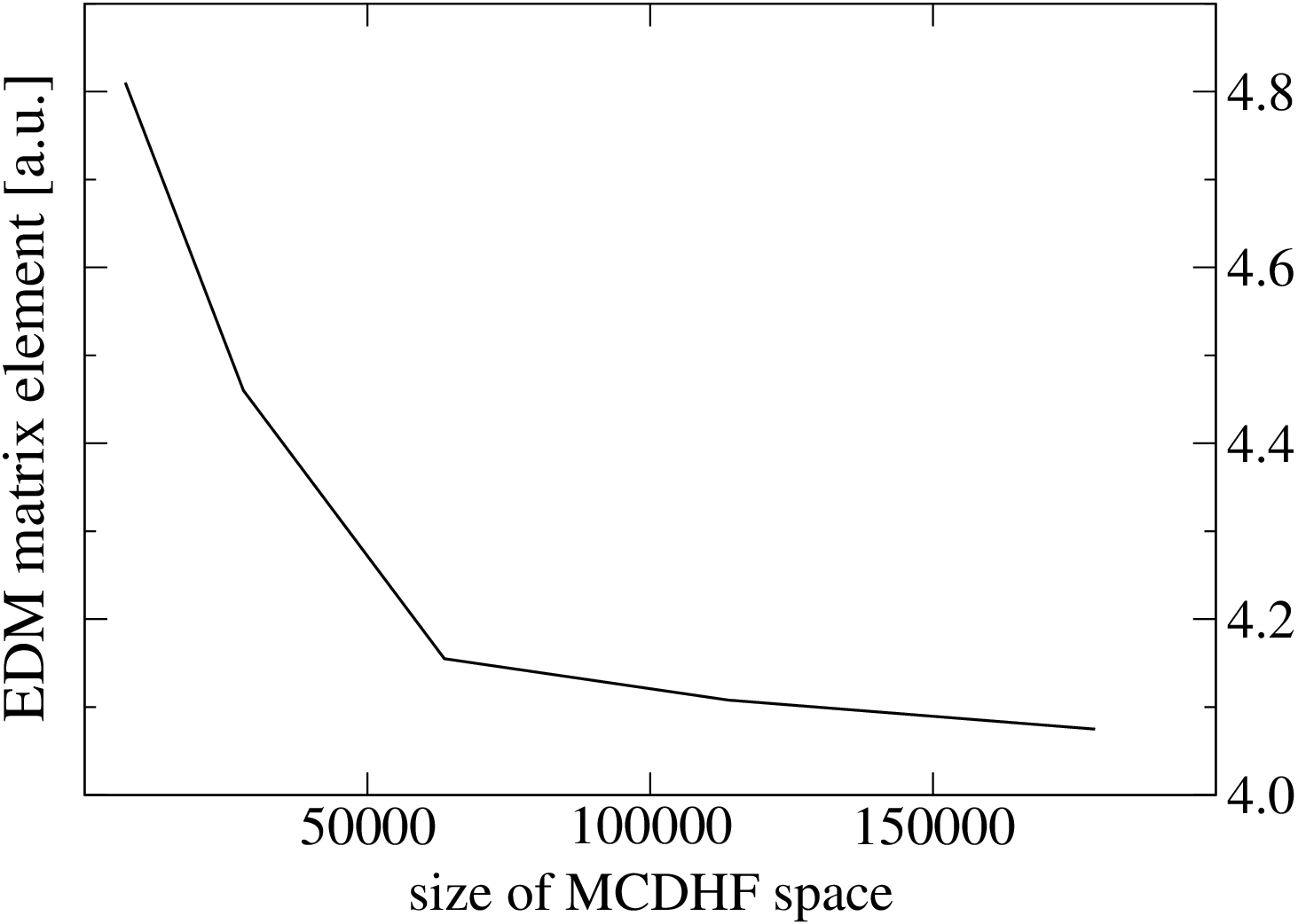}}
    \caption{The values of the reduced matrix elements of the
             (a) Schiff and (b) EDM operators
             as functions of the size of the multiconfiguration expansion}
    \label{figure2_schiff_edm}
\end{figure*}

Table~\ref{Two_matrix_electron_Shiff_element} 
shows the values of the reduced matrix element of the Schiff operators
$4 \sqrt{3} \; \pi \, \left[ {}^3 \! P_1 \left\| \hat{\nabla}^{1} \;
    \rho \left( r  \right) \right\| {}^3 \! D_2 \right]$
from Eq.~(\ref{eq:mat_element_Schiff}).
The expectation values were calculated for the multiconfiguration expansions 
obtained from all possible combinations of virtual sets, and
from opening sequentially the core subshells, as described in 
section~\ref{cic}.

Table~\ref{Two_matrix_electron_EDM_element} presents the values of the 
reduced matrix element of the EDM operator 
$\sqrt{3} \, \left[ {}^3 \! P_1 \left\| -er \right\| {}^3 \! D_2 \right]$
from Eq.~(\ref{eq:mat_element_EDM}).
As in Table~\ref{Two_matrix_electron_Shiff_element},
the EDM matrix elements were calculated for
the multiconfiguration expansions resulting from
all possible combinations of virtual sets and from opening up the
core subshells, cf.~section~\ref{cic}.

%
%
\begin{table}
\caption{The values of the reduced matrix element of the EDM operator
$\sqrt{3} \; \left[ {}^3 \! P_1 \left\| -er \right\| {}^3 \! D_2 \right]$
from Eq.~(\ref{eq:mat_element_EDM}) [a.u.].
The notations are the same as in Table~I.
}
\vspace*{0.3cm}
\label{Two_matrix_electron_EDM_element}
\begin{tabular}{c|ccccccccc}
\colrule  \\[-0.3cm] 
  DF  & 5.849 &  ~~&  ~~&  ~~& ~~&  \\ %
\colrule   \\[-0.3cm] 
\multicolumn{1}{c|}{ virtual } &
\multicolumn{5}{c}{ shells opened for substitutions  } & \\
\multicolumn{1}{c|}{ set } &
 \multicolumn{1}{c}{ ~~~6p~~~  } &
 \multicolumn{1}{c}{ ~~~6sp~~~ } &
 \multicolumn{1}{c}{ ~~5d6sp~~ } &
 \multicolumn{1}{c}{ ~5pd6sp~  } &
 \multicolumn{1}{c}{ ~5spd6sp~ } \\
\colrule   \\[-0.3cm] 
 1v & $4.818$ & $4.811$ & $4.810$ & $4.810$ & $4.810$ \\ %
 2v & $4.514$ & $4.467$ & $4.460$ & $4.460$ & $4.460$ \\ %
 3v & $4.270$ & $4.192$ & $4.157$ & $4.155$ & $4.155$ \\ %
 4v & $4.251$ & $4.168$ & $4.114$ & $4.109$ & $4.108$ \\ %
 5v & $4.240$ & $4.153$ & $4.083$ & $4.076$ & $4.075$ \\ %
\colrule  
%
\end{tabular}
\end{table}
%

The data from Tables~\ref{Two_matrix_electron_Shiff_element} 
and~\ref{Two_matrix_electron_EDM_element}
are collected also in Figure~\ref{figure2_schiff_edm},
where they are presented as functions of the size $\rm NCF$
(see Eq.~\ref{ASF})
of the multiconfiguration expansion.
Figure~\ref{subfigure2_schiff} shows the reduced matrix element of the
Schiff operator,
while Figure~\ref{subfigure2_edm} shows the
value of reduced matrix element of EDM operator.
%
%
The resultant EDM, induced by the nuclear Schiff moment in the
$ ^3 \! D_2 $ state of the  isotope Ra-223
is presented in Table~\ref{EDM_biortogonal_final}.

%
%
%
\begin{table}[b]
\caption{ The atomic EDM $ D_A $  [$ 10^9 S/I $]  of the
 $\ketm{F,M_F} \,=\, \ketm{3/2, 3/2}$ hyperfine
 state of the $ ^3 \! D_2 $ level, induced by the nuclear Schiff moment in the
 isotope ${}_{\hspace{4pt}88}^{223}$Ra ($I \ = \frac{3}{2}$) [a.u.].
 The notations are the same as in Table~I.
}
\vspace*{0.3cm}
\label{EDM_biortogonal_final}
\begin{tabular}{c|ccccccccccc}
\colrule   \\[-0.3cm] 
%
%
 DF & 0.0512 &  ~~&  ~~&  ~~& ~~&  \\ %
\colrule   \\[-0.3cm] 
\multicolumn{1}{c|}{ virtual } &
\multicolumn{5}{c}{ shells opened for substitutions  } & \\
\multicolumn{1}{c|}{ set } &
 \multicolumn{1}{c}{ ~~~6p~~~  } &
 \multicolumn{1}{c}{ ~~~6sp~~~ } &
 \multicolumn{1}{c}{ ~~5d6sp~~ } &
 \multicolumn{1}{c}{ ~5pd6sp~  } &
 \multicolumn{1}{c}{ ~5spd6sp~ } \\
\colrule   \\[-0.3cm] 
 1v & 0.1812 & $0.1181$ ~~& $0.1163$ ~~& $0.1092$ ~~& $0.1028$ \\ %
 2v & 0.2694 & $0.4378$ ~~& $0.4374$ ~~& $0.4399$ ~~& $0.4489$ \\ %
 3v & 0.2850 & $0.4393$ ~~& $0.4378$ ~~& $0.4383$ ~~& $0.4492$ \\ %
 4v & 0.2627 & $0.4293$ ~~& $0.4244$ ~~& $0.4208$ ~~& $0.4329$ \\ %
 5v & 0.2555 & $0.4224$ ~~& $0.4144$ ~~& $0.4097$ ~~& $0.4346$ \\ %
\colrule  
%
\end{tabular}
\end{table}

\subsection{Electron correlation effects}
\label{correlation}

The calculations appear to be converged within the multiconfiguration
approximation employed in the present paper.
A comparison of the 'DF' value with the final result in
Table~\ref{EDM_biortogonal_final} shows, that correlation effects
are dominant in these calculations.
An inspection of Tables~\ref{Two_matrix_electron_Shiff_element}
and ~\ref{Two_matrix_electron_EDM_element}
reveals that the Schiff moment matrix element is mainly responsible
for the correlation correction to the atomic EDM.
Overall, the correlated Schiff value is more than an order of 
magnitude larger than the uncorrelated one.
On the other hand, the correlated value of the EDM matrix element
is about 30\% smaller than the uncorrelated one.
Together, they result in the atomic EDM, which is 8.5 times larger
than the uncorrelated value.
There are several factors, which have to be taken into consideration 
in these calculations in order to capture the bulk of the electron 
correlation effects:
\begin{enumerate}
\item
%
Schiff interaction is localised inside nucleus
(see Eq.~\ref{eq:radial_integral_Schiff}),
therefore only the (one-electron)
matrix elements
$ \left[ s_{1/2} 
\| \hat{\nabla}^{1} \; \rho \left( r \right) \|
p_{1/2} \right] $
and
$ \left[ s_{1/2} 
\| \hat{\nabla}^{1} \; \rho \left( r \right) \|
p_{3/2} \right] $
contribute appreciably in Eq.~(\ref{eq:mat_element_between_CSF})
(see also Table~II in ref.~\cite{Dzuba2000}).
All other spin-orbitals may contribute
through the indirect electron correlation effects~\cite{Bieron:Au:2008}.
\vspace*{-0.25cm}
%
\item
Only the inner regions of the spin-orbitals contribute
to the Schiff moment matrix elements.
Therefore it is essential to accurately represent
core polarisation in the atomic wave functions.
\vspace*{-0.25cm}
\item
For the electric dipole interaction, the outer parts of the spin-orbitals
 are important.
Therefore the valence correlation effects have to be accounted for.
%
%
The Babushkin and Coulomb gauge values differ by 5 orders of magnitude
in the calculation of the very weak
$ ^3 \! P _1 - \, ^3 \! D _2 $ 
transition~\cite{Bieron:Ratau:2007}.
This results from the fact, that the transition energy is very small.
Transition energy is difficult to reproduce accurately
in variational calculations, when the wave functions for the two levels
are generated separately and when their one-electron orbitals are 
optimised independently.
It is necessary to obtain a balanced description of both
states~\cite{Fischer:1999}, which can only be achieved if all
two-body (double substitutions) and 
possibly also three- and four-body (triple and quadruple substitutions)
correlation effects are fully taken into account.
%
In general, the two gauges exhibit different energy dependence:
the Coulomb gauge is strongly energy dependent, therefore the
Babushkin gauge value is usually adopted in unsaturated calculations,
since it is less dependent on calculated transition energy value.
The calculations of energy level differences require well balanced 
orbital sets and typically demand highly
extensive multiconfiguration expansions. 
The results are usually in better agreement with experiment if a common 
set of orbitals is used for both states.
\vspace*{-0.25cm}
\item
On the other hand, the wave functions for the two levels have to be 
generated separately, if the effects of non-orthogonality are to be 
taken into account correctly. Non-orthogonality between the
spin-orbitals is essential in the evaluation of non-diagonal matrix elements 
(see Ref.~\cite{Bieron:Ratau:2007}.
\end{enumerate}

\subsection{Breit and QED corrections}
\label{breit-qed}

The calculations descibed in the previous sections
as well as the data in
Tables~\ref{Two_matrix_electron_Shiff_element},%
~\ref{Two_matrix_electron_EDM_element}, and
\ref{EDM_biortogonal_final}
were obtained with
only the Coulomb interaction included
in the differential equations which are iteratively solved during
the self-consistent field process (SCF).
The effects of magnetic and retardation corrections can be evaluated by
introducing the full Breit operator in the self-consistent field process.
Such modified differential equations lead to change
of the final wavefunction shape, which in turn modifies the matrix elements
evaluated in the calculation of the Schiff moment.
Quantum electrodynamics (QED) correction is dominated by two contributions.
The self-energy (SE) part cannot be easily evaluated.
Its contribution to the Cs parity violation amplitude has been calculated
in Ref.~\cite{spty2005} and turned out to be of the order of -0.7~\%.
On the other hand, the vacuum polarization can be easily calculated by
introducing the Uelhing potential into the SCF, as was
done, for example, in calculations of Li-like ions hyperfine matrix
elements~\cite{bai2000} or in Cs parity-violation amplitude~\cite{jbs2001}.
In the latter case, it leads to a 0.4~\% increase of the amplitude.
If both are evaluated, the SE and VP contributions partly compensate.
We have used the \emph{mdfgme} code~\cite{mcdfgme} in its 2008 version,
which now includes a Schiff moment option, to evaluate the effect of
the Breit interaction and vacuum polarization on the Schiff moment.
The results are presented in Table~\ref{tab:bsc-vp}.
The inclusion of the Breit interaction in the SCF reduced the value
of the Schiff moment by a factor of 1.7~\%,
while the vacuum polarization increased it by roughly 0.6~\%.
The total correction is a reduction of the order of 1.1~\%.
From Ref.~\cite{spty2005} one should expect a partial, mutual compensation
of the VP and SE contributions.
The total correction would then be dominated by the Breit contribution.

The results presented in Table~\ref{tab:bsc-vp} were obtained in
single configuration approximation, i.e., without electron correlation
effects.
Our experience indicates, that a fully
correlated calculation, i.e., 
with the multiconfiguration expansion~(\ref{ASF})
described in section~\ref{cic},
and with the self-consistent Breit interaction  would
lead to a larger (absolute value of) total correction.
However, the increase of the number of extra integrals involved
(roughly by 2 orders of magnitude)
and substantial convergence difficulties
render such a calculation virtually impossible with today's computers. 
 
%
%
The final correction from 
Table~\ref{tab:bsc-vp}
(i.e., the reduction by a factor of 0.989~\%)
 has been carried over to the final summary
of our results, which is presented in 
Table~\ref{EDM_biorthogonal_3iso}.

\begin{table}[tb]
  \centering
  \begin{ruledtabular}
  \begin{tabular}{lrrrr}
& \multicolumn{1}{c}{ Schiff }
                  & \multicolumn{1}{c}{ Stark }
                                  & \multicolumn{1}{c}{ EDM }
                                          & \multicolumn{1}{c}{ correction }\\
  \hline
DF      	&	674	&	5.849	&	0.0512	&	\\
  \hline								
Breit SCF            &	664	&	5.835	&	0.0503	& $-$1.7\;\% \\
  \hline								
Breit SCF + VP SCF &    668	&	5.837	&	0.0507	& $-$1.1\,\% \\ 
  \end{tabular}
  \end{ruledtabular}  
\caption{Effect of the Breit and vacuum polarization on the
  reduced matrix element of the Schiff operator (Schiff),
  reduced matrix element of the EDM operator (Stark),
  and
  the atomic EDM ($ D_A $  [$ 10^9 S/I $])  of the
 $\ketm{F,M_F} \,=\, \ketm{3/2, 3/2}$ hyperfine
 state of the $ ^3 \! D_2 $ level, induced by the nuclear Schiff moment in
 the isotope ${}_{\hspace{4pt}88}^{223}$Ra ($I \ = \frac{3}{2}$) [a.u.].
VP --- vacuum polarization.
DF --- Coulomb interaction only.
}
\label{tab:bsc-vp}
\end{table}

\subsection{Accuracy}
\label{accuracy}

Although we have full control of the core polarisation effects,
the overall accuracy of these results depends primarily on the
electron-correlation effects which were not included in the calculations,
i.e.~unrestricted double substitutions and triple substitutions.
These unrestricted substitutions had to be omitted
due to software and hardware limitations.
A similar approximation,
based on single and restricted double substitutions, was employed also
in our previous papers on radium~\cite{%
Bieron:Ra3d2:2004,%
Bieron:RaQ:2005,%
Bieron:Rahfs:2005,%
Bieron:Ratau:2007}.
The accuracy of the present calculations can be indirectly inferred
from a comparison of the previously calculated transition rates and
hyperfine constants with experiment.
Reference~\cite{Bieron:Rahfs:2005} showed hyperfine constants
calculated in an approach similar to that employed in the present paper.
As discussed there, the overall accuracy of the calculated magnetic dipole 
constants was 6~\%{}, while for the electric quadrupole constants the 
estimated accuracy was 3~\%{}.
The interaction responsible for the hyperfine shifts 
takes place in the vicinity of the nucleus. 
The bulk of 
the hyperfine integral/matrix elements come from the first oscillation 
of the one-electron wave function~\cite{Pyykko:1972,Pyykko:1973}.
Therefore,
the calculated hyperfine constants depend heavily on 
the inner regions of the one-electron wave functions.
The interaction of the electronic cloud with the nuclear Schiff moment
also takes place in the vicinity of the nucleus. In fact, the bulk of the
Schiff integral/matrix~element comes from the nuclear
skin~(see Eq.~\ref{eq:radial_integral_Schiff}). Therefore we might 
expect that the accuracy of the calculated Schiff moment matrix 
elements is comparable to the accuracy of the calculated hyperfine 
constants.
In conclusion, we might expect the accuracy of the calculated 
Schiff moment matrix elements of the order of 10~\% or better.

The electric-dipole moment matrix 
element~(see~Eq.~\ref{eq:radial_integral_EDM})
has similar radial dependence as the electric-dipole transition elements,
therefore their accuracy might also be expected to be of the same order.
The calculated rate for the strong transition 
$ ^1 \! P _1 \, $---$ \, ^1 \! S _0 $ showed excellent agreement between the 
values calculated in Babushkin and Coulomb gauges~\cite{Bieron:Ra3d2:2004},
which is an indication (but not a proof) of convergence.
The accuracy of the calculated rate for the
$ ^3 \! P _1 \, $---$ \, ^1 \! S _0 $
transition was also reasonably good~\cite{Bieron:Ratau:2007}.
Although there was a 20~\%{} difference between the values calculated
in the two gauges, the Babushkin gauge value fell within the 
experimental limit~\cite{Scielzo2006}.
The gauge difference was somewhat larger (35~\%) for the calculated rate 
for the
$ ^3 \! D _2 \, $---$ \, ^1 \! S _0 $ electric quadrupole
transition~\cite{Bieron:Ra3d2:2004} which is the only significant 
decay channel of the metastable state $ ^3 \! D _2 $.
However, the calculated rates of other weak transitions were less accurate.
In particular, the calculated rate for the
$ ^3 \! P _1 \, $---$ \, ^3 \! D _2 $
transition showed very large gauge dependence, which seems to indicate
that the core-core correlation effects (which were omitted in present 
calculations) are important
(see e.g.~\cite{Joensson:Na:1996}).
These correlation contributions are currently beyond the capacity 
of the computer resources available to us.
We were unable to include, or even to estimate the contributions of
the omitted electron correlation effects,
i.e.,~the unrestricted double substitutions and triple substitutions,
to the atomic EDM.
Such calculation would require a multiprocessor cluster in excess
of a hundred processors~\cite{Bieron:AuCAS:2009}.
In order to estimate the uncertainty of the electric dipole moment
 calculation, we had to resort to an upper limit resulting from
the hierarchy of electron-electron interactions.
Table~\ref{Two_matrix_electron_EDM_element}
demonstrates the saturation of the 
core-valence correlation correction, which is
the dominant electron correlation effect beyond 
the  Dirac-Fock approximation.
Therefore we might expect that the entire (core-valence) electron 
correlation contribution would be a rather conservative estimate
of the uncertainty of the electric dipole moment calculations.
%
%
In conclusion, we may assume the accuracy of the calculated 
electric dipole moment matrix elements of the order of 30~\%.
%

Another contribution that affects the overall accuracy of the results
arises from the nuclear-density dependence of radial matrix elements 
in Eq.~(\ref{eq:radial_integral_Schiff}), i.e.~from the
distribution of the nuclear charge, as discussed in 
section~\ref{grid}.
%
However, the error of the calculated value of atomic EDM is 
dominated by the electric dipole moment calculation.
%
Therefore we assumed 30~\% accuracy of the final calculated 
value of the Schiff moment enhancement factor,
as shown in Table~\ref{EDM_biorthogonal_3iso}.

A more stringent accuracy assessment for the present
calculations would require more experimental data to compare with.
Hereby, we would like to encourage the experimenters to provide them.
In particular, hyperfine constants for other levels, as well as
lifetime and transition rate measurements for weak transitions would be 
very valuable.
 
The results presented in the second and in the last column of
 Table~\ref{EDM_biorthogonal_3iso} illustrate the isotope-dependence
of the atomic EDM. The two isotopes in question, 
$ ^{213}_{\phantom{1}88}$Ra 
and
$ ^{225}_{\phantom{1}88}$Ra,
have the same nuclear spin ($I = 1/2$), but they
have slightly different nuclear shapes
 (see Eq.~(\ref{two-parameter-fermi.eq})
  and the subsequent discussion in section~\ref{grid}).
The abovementioned isotope-dependence, of the order of 0.4~\%,
arises primarily through the derivative 
 $ {d \rho} / {dr} $
in Eq.~(\ref{eq:radial_integral_Schiff}).
The isotope-dependence of the atomic wave functions~(\ref{ASF})
has been neglected in the present calculations.

%
\begin{table}
\caption{ The atomic EDM induced by the nuclear Schiff moment
in the $ ^3 \! D_{2} $ electronic state 
for three isotopes of radium:
 $ ^{213}_{\hspace{4pt}88}$Ra~($ I = \frac{1}{2}, F = \frac{3}{2}$),
 $ ^{223}_{\hspace{4pt}88}$Ra~($ I = \frac{3}{2}, F = \frac{3}{2}$), and
 $ ^{225}_{\hspace{4pt}88}$Ra~($ I = \frac{1}{2}, F = \frac{3}{2}$),
respectively. Results are shown as functions of the size of the
virtual orbital set, including electron substitutions from the
spectroscopic $5spd6sp[7s6d|7s7p] $ orbitals to:
 (1v) one layer of virtual orbitals,
 (2v) two layers of virtual orbitals, ... ,
 (5v) five layers of virtual orbitals.
'DF' = uncorrelated Dirac-Fock value.
%
%
Our values include the 'Breit SCF + VP SCF' correction
from Table~\ref{tab:bsc-vp}.
The RHF+CI results in the last line are quoted from Ref.~\cite{Dzuba2000}.}
%
\vspace*{0.3cm}
\label{EDM_biorthogonal_3iso}
\begin{tabular}{l|d|d|d}
\colrule  \\[-0.3cm]  
 & \multicolumn{3}{c}{ EDM [$10^9 S/I$] [a.u.] } \\[0.1cm]
\cline{2-4}
& & & \\ [-0.3cm]
layer
&
\multicolumn{1}{c|}{${}^{213}_{\phantom{1}88}$Ra}
&
\multicolumn{1}{c|}{${}^{223}_{\phantom{1}88}$Ra}
&
\multicolumn{1}{c }{${}^{225}_{\phantom{1}88}$Ra}
\\
\colrule  
& & & \\[-0.3cm]
 DF & 0.0159 & 0.0507 & 0.0158  \\
 1v & 0.0320 & 0.1017 & 0.0318  \\
 2v & 0.1393 & 0.4441 & 0.1387  \\
 3v & 0.1394 & 0.4444 & 0.1388  \\
 4v & 0.1344 & 0.4283 & 0.1338  \\
 5v & 0.1349 & 0.4300 & 0.1343  \\[0.05cm]
\colrule
& & & \\[-0.3cm]
      final  & 0.13(4) & 0.43(14) & 0.13(4) \\
\colrule
& & & \\[-0.3cm]
RHF+CI & 0.094 & 0.30 & 0.094   \\
\colrule  
%
\end{tabular}
\end{table}

\section{Conclusions}
\label{conclusions}

Radium is well suited for EDM experiments because of its large
nuclear charge $Z$ and the two low-lying levels
 $ ^3 \! P_1 $ and $ ^3 \! D_2 $ of opposite
parity, which are separated in energy by only
5~cm$^{-1} \;\approx 2 × 10^{-5}$ a.u.
The $ ^3 \! D_2 $ level is metastable with a
very long lifetime
of the order of 4~seconds~\cite{Bieron:Ratau:2007}
and therefore suitable for
laser-ion traps.
For radium, moreover, there are several isotopes with mass ranging
from 209 up to 229, of which
$ ^{213}_{\phantom{1}88}$Ra 
and 
$ ^{225}_{\phantom{1}88}$Ra 
 have nuclear spin $I=1/2$. For EDM experiments the $I=1/2$ isotopes
are preferable since these isotopes cannot be disturbed by
higher order electromagnetic moments.
%

The paper presents
systematic computations of the static EDM of atomic 
radium in the $ ^3 \! D_2 $ level, induced by the nuclear Schiff moment. 
Table~\ref{EDM_biorthogonal_3iso} shows the 
atomic EDM induced by the nuclear Schiff moment
in the $ ^3 \! D_{2} $ electronic state
for three isotopes of radium:
\begin{itemize}
\item[]
\vspace{-8pt}
 $ ^{213}_{\hspace{4pt}88}$Ra~($ I = \frac{1}{2}, F = \frac{3}{2},
  \mu = 0.6133$),
\item[]
\vspace{-8pt}
 $ ^{223}_{\hspace{4pt}88}$Ra~($ I = \frac{3}{2}, F = \frac{3}{2},
  \mu = 0.2705$), and
\item[]
\vspace{-8pt}
 $ ^{225}_{\hspace{4pt}88}$Ra~($ I = \frac{1}{2}, F = \frac{3}{2},
  \mu = -0.7338$).
\end{itemize}
\vspace{-8pt}
The wave functions for the two levels,
 $ ^3 \! P_1 $ and $ ^3 \! D_2 $, were
generated separately,
in order to correctly reproduce
the effects of non-orthogonality between one-electron spinorbitals.
We demonstrate that core-valence electron correlation,
which is the dominant
electron correlation effect beyond the Dirac-Fock approximation,
contributes almost 90\% of the total EDM value.
Our final value is about 30\% larger than 
the RHF+CI result
from Ref.~\cite{Dzuba2000}.
%
The difference can be atributed to different methods
employed to account for core-valence electron correlation effects.
Neither of these two calculations included the core-core
correlation effects, which we believe to be the dominant source
of uncertainty in our calculated value of the Schiff moment
enhancement factor.

\section*{Acknowledgements}
\noindent
This work was supported by
the Polish Ministry of Science and Higher Education (MNiSW)
in the framework of the scientific grant No.~1~P03B~110~30
awarded for the years 2006-2009.
SF acknowledges the support by the DFG under the project No.~FR~1251/13.
EG acknowledges the Student Research Fellowship Award from the
Lithuanian Science Council.
PJ acknowledges the support from the
Swedish Research Council (Vetenskapsr{å}det). 
PI acknowledges the support of the
Helmholtz Alliance Program of the Helmholtz Association, contract
HA-216 ``Extremes of Density and Temperature: Cosmic Matter in the
  Laboratory''.
Laboratoire Kastler Brossel is ``Unit{é} Mixte de Recherche du CNRS, de
l'ENS et de l'UPMC n$^{\circ}$ 8552''.
We would like to thank Klaus Jungmann, Jeffrey Guest, William Trimble
and Jean-Paul Desclaux
for helpful discussions.


\bibliography{xet}
\end{document}